# Elastically Cooperative Activated Barrier Hopping Theory of Relaxation in Viscous Fluids. II. Thermal Liquids

Stephen Mirigian and Kenneth S. Schweizer*

Departments of Material Science and Chemistry, and Frederick Seitz Materials Research Laboratory, University of Illinois, 1304 W. Green Street, Urbana, IL 61801
*kschweiz@illinois.edu

## Abstract

Building on the elastically collective nonlinear Langevin equation theory developed for hard spheres in the preceding paper I, we propose and implement a quasi-universal theory for the alpha relaxation of thermal liquids based on mapping them to an effective hard sphere fluid via the dimensionless compressibility. The result is a zero adjustable parameter theory that can quantitatively address in a unified manner the alpha relaxation time over 14 or more decades. The theory has no singularities above zero Kelvin, and relaxation in the equilibrium low temperature limit is predicted to be of a roughly Arrhenius form. The two-barrier (local cage and long range collective elastic) description results in a rich dynamic behavior including apparent Arrhenius, narrow crossover and deeply supercooled regimes, and multiple characteristic or crossover times and temperatures of clear physical meaning. Application of the theory to nonpolar molecules, alcohols, rare gases and liquids metals is carried out. Overall, the agreement with experiment is quite good for the temperature dependence of the alpha time, plateau shear modulus and Boson-like peak frequency for van der Waals liquids, though less so for hydrogen-bonding molecules. The theory predicts multiple growing length scales upon cooling, which reflect distinct aspects of the coupled local hopping and cooperative elastic physics. Calculations of an activation volume that grows with cooling, which is correlated with a measure of dynamic cooperativity, agree quantitatively with experiment. Comparisons with elastic, entropy crisis, dynamic facilitation and other approaches are performed, and a fundamental basis for empirically-extracted crossover temperatures is established. The present work sets the stage for addressing distinctive glassy phenomena in polymer melts, and diverse liquids under strong confinement.

## I. INTRODUCTION

The problem of slow dynamics in supercooled liquids remains a grand challenge of soft condensed matter science. Inherent to this problem is multiple temperature "regimes", characteristic (crossover and extrapolated) temperatures, and energy scales [1-5]. What aspects of this rich phenomenology are "fundamental" versus "apparent" features associated with empirical fitting remains vigorously debated. Creating a microscopic physical basis for such complexity over 14 or more orders of magnitude of relaxation time is challenging. We believe definitive progress requires a unified quantitative description of all dynamical regimes. In this article, we address this problem based on the Elastically Cooperative Nonlinear Langevin Equation (ECNLE) approach developed in the preceding paper I [6] and propose a theory that can serve as a *zeroth order* quasi-universal description of thermal liquids.

Our strategy is based on a "mapping" of real molecules to an effective hard sphere fluid guided by requiring the latter exactly reproduces the "long" (determined in practice on the ~nm scale) wavelength equilibrium *dimensionless* density fluctuation amplitude of a liquid, $S_0$, a well-defined thermodynamic property [7]. This mapping yields a system-specific and thermodynamic-state-dependent effective hard sphere volume fraction, $\phi$, which encodes in an averaged sense the thermodynamic consequences of repulsive and attractive forces and molecular shape. The resultant theory connects thermodynamics, structure and dynamics in the simplest manner we can envision for a force-level approach. Although there are limitations, a large advantage is that a priori predictions can be made that are genuinely falsifiable since no adjustable/fit parameters enter the theory. We anticipate the mapping formulated here will be most useful for van der Waals (vdW) liquids ("strongly correlating" liquids of the Roskilde group [8-10]), and less accurate as chemical/structural complexity is introduced, e.g., hydrogen-bonding, ionic interactions, network formers.

Section II presents our mapping and develops several of its general consequences. Limiting analytic results for key length and energy scales and short and long time properties in the deeply supercooled regime are discussed in section III; an analysis of the equilibrium low temperature limit is also presented. Section IV presents representative numerical calculations and comparisons to experiments for the alpha relaxation time, $T_g$, fragility, shear modulus, and characteristic vibrational frequency for 12 glass forming liquids including nonpolar molecules, alcohols, rare gases and liquid metals; important crossover temperatures and time scales are also discussed. The similarities and differences between our approach and the phenomenological shoving model [11, 12] are established in section V, including the relative role of the shear modulus and a growing dynamical length scale in determining the collective barrier in the deeply supercooled regime. Section VI presents calculations for the effect of pressure on the alpha relaxation, and also analyzes an "activation volume" that grows with cooling and is strongly correlated with the number of cooperatively moving molecules and other measures of dynamic heterogeneity [13, 14]. Connections of our approach with diverse alternative theories and models, including Arrhenius, mode coupling theory (MCT) [15], entropy crisis [16, 17], dynamic facilitation [18], and phenomenological two-barrier approaches [19-22] is the subject of section VII. Our theoretical results are treated as "data" and we explore how well these models can fit our calculations. This exercise also allows the empirical extraction of characteristic temperatures and time scales, and their physical meaning to be deduced. The article concludes in section VIII with a discussion. For economy of expression, we assume the reader is familiar with the preceding paper I [6], and equations from that article are cited as Eq(I.x).

## II. MAPPING TO THERMAL LIQUIDS AND QUASI-UNIVERSALITY

### A. Density Fluctuations

The thermodynamic state and material-dependent dimensionless amplitude of density fluctuations is determined by the molecular number density, thermal energy and isothermal compressibility, or alternatively as a specific derivative of pressure, as [7]

$$S_0^{\text{expt}} = \rho k_B T \kappa_T = \left(\frac{\partial \beta P}{\partial \rho}\right)^{-1} \quad (1)$$

This quantifies the "flat" part of the structure factor, $S(k)$, at low wavevectors which emerges in practice on scales beyond the local (typically nm) structural correlations in liquids. The mapping then corresponds to enforcing the equality

$$S_0^{\text{expt}} \equiv S_0^{HS}(\phi) \quad (2)$$

which defines $\phi$ of the reference hard sphere fluid from the liquid equation-of-state (EOS). Eq.(2) corresponds to a quasi-universal picture where the dynamics of all liquids follow from a hard sphere fluid to within the nonuniversal prefactor in Eq(I.35) associated with binary collision physics [23, 24]. No separation of $\phi$ into a number density and hard sphere diameter is required. The mapping effectively replaces the volume fraction axis of the alpha time plots in paper I by temperature in a system-specific and thermodynamic state dependent manner.

We expect this mapping idea to work best for nearly spherical vdW-like molecules, and for the longer range (more coarse-grained) collective dynamics. Based on paper I, we do not believe athermal (particle-shape-dependent) jamming is important for equilibrated liquids. Given this, and the simplicity of replacing real molecules with spheres, we employ the simplest integral equation for all equilibrium quantities, the compressibility route Percus-Yevick (PY) theory [7].

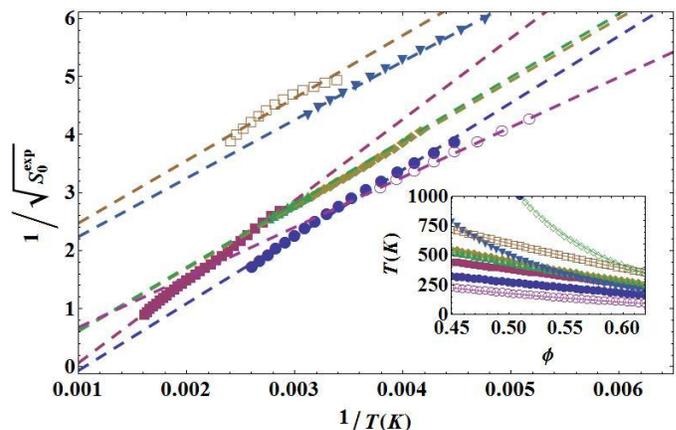

**Figure 1**: Experimental dimensionless compressibility data for toluene(blue circles), biphenyl(red squares), OTP(yellow diamonds), salol(green upward triangles – obscured by OTP), glycerol(gray downward triangles), ethanol(open red circles), and sorbitol(open orange squares). The dashed lines of corresponding color are fits to Eq. 8. Inset: The resulting mapping, Eq. (11), between volume fraction and temperature for toluene(blue circles), biphenyl(red squares), OTP(yellow diamonds), salol(green upward triangles – obscured by OTP), glycerol(gray downward triangles), ethanol(open red circles), TNB(open yellow squares) and sorbitol(open diamonds). In the absence of equation of state data, we have taken the A and B parameter values for TNB to be the same as OTP.

Fig. 1 presents experimental dimensionless compressibility data for diverse liquids. For some systems, the full EOS data is not available at either very high and/or very low temperatures, and thus Eq.(1) cannot be used directly and extrapolation is required. Motivated by this practical point, and also the desire for physical insight into the material-dependence of the dimensionless compressibility, we perform an analytic analysis of the classic vdW EOS.

**B. van der Waals Analysis of Dimensionless Compressibility**

The vdW model expresses the pressure as additive contributions of repulsive and attractive (cohesive) interactions [7]:

$$\beta P = \beta P_{rep} + \beta P_{att} = \frac{\rho}{1-b\rho} - a\rho^2 \quad (3)$$

where $a$ and $b$ quantify the integrated strength of the intermolecular attraction in units of the thermal energy (mean field cohesive energy) and molecular volume, respectively. The inverse dimensionless compressibility is

$$S_0^{-1} = \frac{1}{(1-b\rho)^2} - 2a\rho \quad (4)$$

Atmospheric pressure conditions are well approximated by taking P→0, whence one obtains,

$$S_0^{-1} = a\rho(a\rho - 2) \quad (5)$$

$$\rho = \frac{1}{2b}\left(1 + \sqrt{1 - \frac{4b}{a}}\right) \quad (6)$$

For b/a<<1, Eq.(4) then simplifies to:

$$S_0^{-1} \approx \frac{a^2}{b^2} - 4\frac{a}{b} + 4 + ... \approx \left(\frac{a}{b} - 2\right)^2 \quad (7)$$

suggesting the simple analytic form:

$$\frac{1}{\sqrt{S_0^{expt}}} \approx \frac{a}{b} - 2 \equiv \frac{B'}{T} - A' \quad (8)$$

The defined parameters B' and A' are the molecular level liquid cohesion and an entropic or packing contribution. Eq.(8) was the motivation for the plot format in Fig. 1.

Though *not* necessary to implement our mapping, to explicitly reveal the distinct dynamical consequences of molecular size and "intrinsic" chemical effects we imagine the molecule consists of $N_s$ rigidity bonded "interaction sites" (e.g., a site in benzene is a CH group). The dimensionless compressibility can then be written at the site level as:

$$S_0^{site} = \rho_s k_B T \kappa_T \equiv N_s \rho k_B T \kappa_T = N_s S_0^{expt} \quad (9)$$

Eqs.(8) and (9) imply the relation between the molecular and site level dimensionless compressibility parameters is: $A = A'/\sqrt{N_s}$ and $B = B'/\sqrt{N_s}$.

Extracted values of A and B (and $N_s$ values) are given in Table 1 for four classes of substances: 2 rare gases, 2 metals, 5 non-polar vdW molecules, and 3 alcohols; EOS data for TNB was not available and OTP parameters were used. Despite the crude basis of Eq.(8), it does a surprisingly good job of linearizing experimental data even for ethanol and glycerol. However, the sorbitol data is not well linearized; while we report the fit values of A and B for it, a more reliable approach is to directly use Eq.(2) and results based on it are denoted with an asterisk in the Tables. For all the other systems studied, differences in our dynamical predictions based on using Eqs.(2) and (8) are negligible.

Several interesting chemical trends are evident in Table 1. The rare gases and vdW molecules have similar B values, while the alcohols (metals) have smaller (larger) values reflecting their different intermolecular attractions. The "entropic packing" factor is more variable, with positive values for rare gases commensurate with the literal vdW model value of A=2. Smaller positive values are found for the vdW molecules, and even smaller and/or negative values for hydrogen-bonders and metals. Physically, as A decreases, the rate at which the thermal density fluctuation amplitude decreases with cooling is reduced, suggestive of a structurally "stronger" liquid, an intuitive trend.

**C. Analytic Implementation of Quasi-Universal Description**

Using the analytic compressibility route PY theory expression for $S_0$, one has[7]

$$S_0^{HS} = \frac{(1-\phi)^4}{(1+2\phi)^2} \equiv N_s^{-1}\left(-A + \frac{B}{T}\right)^{-2} \quad (10)$$

Solving for the effective volume fraction and employing Eq. 2 yields

$$\phi(T; A, B, N_s) = 1 + \sqrt{S_0^{expt}(T)} - \sqrt{S_0^{expt}(T) + 3\sqrt{S_0^{expt}(T)}} \quad (11)$$

By using Eqs. (8) and (9), an explicit dependence of ϕ on $A$, $B$ and $N_s$ can be written. The inset of Fig.1 shows calculations of ϕ(T). The 1-to-1 mapping between temperature and ϕ or $S_0$ of the reference hard sphere fluid provided by Eq.(11) can be inverted using Eqs. (8) and (9) to give

$$T(\phi) = \frac{B}{A + \frac{1}{\sqrt{N_s S_0^{HS}(\phi)}}} \quad (12)$$

This relation in conjunction with ECNLE theory provides a no adjustable parameter prescription for calculating the alpha time of any material for which EOS data is available. All the characteristic/crossover volume fractions of the hard sphere fluid discussed in paper I translate to characteristic temperatures, and any characteristic temperature ratio is:

$$\frac{T_2}{T_1} = \frac{A\sqrt{N_s} + S_0^{-1/2}(\phi_1)}{A\sqrt{N_s} + S_0^{-1/2}(\phi_2)} \quad (13)$$

The cohesive energy parameter, B, sets an energy scale for $T_g$, but cancels out in ratios. This has many implications, e.g., characteristic temperature ratios become closer as molecular size and/or packing parameter (A) increase, trends which will be shown correlate with enhanced fragility.

Independent of the dynamic theory, the mapping predicts a simple approximate relation for $T_g$ of chemically homologous molecules (same *A* and *B*). Since $S_0(\phi_g \approx 0.615) \approx 0.0044$, and given the typical A and $N_s$ values in Table 1, to a good approximation $A\sqrt{N_s S_0(\phi_g)} \ll 1$, and hence from Eq.(12) one obtains $T_g \propto B\sqrt{N_s}$. Thus, $T_g$ scales essentially as the square root of the molecular mass, an intriguing trend that has been experimentally established recently for several homologous series [25]. For the homologous pair OTP and TNB in Table 1, $T_g$ is 246K and 346K, which obeys essentially exactly the square root law.

Implicit to our mapping is an assumed underlying universality whereby all thermal liquid relaxation time data would, to zeroth order, collapse when plotted against the dimensionless compressibility. Such a plot is shown in Fig. 2. By construction, theory curves (computed as described below) collapse perfectly up to a material-specific short time scale associated with the collision time prefactor in Eq(I.35). The collapse of experimental data is of course imperfect, and not unexpected given the modeling and statistical mechanical approximations, but nonetheless we believe significant and encouraging.

## III. GENERAL ASPECTS AND LIMITING ANALYTIC ANALYSES

From paper I, the mean alpha relaxation time is [26, 27]:

$$\frac{\tau_\alpha}{\tau_s} = 1 + \frac{2\pi}{\sqrt{K_0 K_B}} \exp\left(\frac{F_B + F_{elastic}}{k_B T}\right) \quad (14)$$

where the "short time" is

$$\tau_s = \frac{g(d)d}{24\phi} \sqrt{\frac{\pi M}{k_B T}} \times \left[1 + \frac{1}{36\pi\phi} \int_0^\infty dQ Q^2 \frac{(S(Q)-1)^2}{S(Q)+b(Q)}\right] \quad (15)$$

$g(d) = (1+\phi/2)/(1-\phi)^2$ and S(Q) is the structure factor with Q=kd. The short time scale contains the only source of explicit nonuniversality in the dynamical theory based on the present minimalist mapping, and is proportional to the inverse Enskog binary collision rate [7], $\tau_E^{-1} = 24\phi g_d d^{-1} \sqrt{k_B T / \pi M}$, which depends on temperature and the molecular diameter and mass. This nonuniversal variation is weak, typically increasing only by ~2.5 upon cooling over the wide temperature range studied here; $\tau_E$ can reasonably be taken to be a constant of ~0.1 psec.

All alpha times are numerically calculated as described in paper I. However, limiting analytic results were also derived in paper I, and it is of interest to first examine their mathematical form based on the thermal mapping in the deeply supercooled regime.

### A. Energy and Length Scales

From section IIIC of paper I [6], the localization length is proportional to the dimensionless compressibility in the deeply supercooled regime, allowing us to write:

$$r_{loc}/d \approx \frac{15}{4} S_0^{HS} \propto N_s^{-1}(-A+B/T)^{-2} \quad (16)$$

All other quantities in the theory, such as the barrier location, the local barrier, and the collective elastic barrier can related to the localization length. The barrier position is:

$$r_B \approx \frac{r_{cage}}{2\pi} \sqrt{3\ln\left(\sqrt{\frac{3}{\pi}}\frac{1}{r_{loc}}\right)}$$
$$\propto \sqrt{const + 6\ln\left[\sqrt{N_s}\left(\frac{B}{T}-A\right)\right]} \quad (17)$$

where the last proportionality assumes the location of the first minimum of g(r), $r_{cage}$, is independent of temperature. The local cage NLE barrier is given by Eq(I.24) as:

$$\beta F_B \approx 0.45 \frac{r_{cage}}{r_{loc}} \frac{3\sqrt{3\pi}}{2\pi^2}$$
$$\propto S_0^{-1} \propto N_s \left(\frac{B}{T}-A\right)^2 \quad (18)$$

In the ultra-local limit, the collective elastic barrier is given by Eq. (I.25). Since in the deeply supercooled regime the localization length is very small, the jump length is essentially equal to $r_B$, and thus one has to good approximation

$$\beta F_{elastic} \propto \phi(T) N_s^2 \left(-A+\frac{B}{T}\right)^4 \times \left(const + 6\ln\left[\sqrt{N_s}\left(\frac{B}{T}-A\right)\right]\right)^2 \quad (19)$$

It was also shown in Paper I (see Eq(I.22)) that in the deeply supercooled regime the local and collective barriers are related to a very good approximation as $F_{elastic} \propto F_B^2$; this relation also follows from Eqs. (18) and (19) by neglecting the logarithmic term and the weak temperature dependence of $\phi(T)$. The total barrier is then:

$$F_{total} = F_B(1+bF_B)$$
$$\propto N_s\left(\frac{B}{T}-A\right)^2 \times \left(1+c\cdot N_s\left(\frac{B}{T}-A\right)^2\right) \quad (20)$$

where $c$ is a constant. This temperature dependence does not correspond to any model or theory we are aware.

The dynamic plateau shear modulus is

$$G = \frac{k_B T}{60\pi^2} \int_0^\infty dk \left[k^2 \frac{d}{dk}\ln(S(k))\right]^2 \exp\left(-\frac{k^2 r_{loc}^2}{3S(k)}\right)$$
$$\propto \phi \frac{k_B T}{d r_{loc}^2} \propto \frac{k_B T}{d^3} N_s(-A+B/T)^2 \quad (21)$$

where the final proportionality again neglects the weak temperature dependence of $\phi(T)$. Knowledge of the cooperative elastic barrier and shear modulus defines the "cooperative volume" of paper I: $V_c(T) \equiv F_{elastic}(T)/G(T) \propto (r_B - r_{loc})^4/d \equiv \Delta r^4/d$. This quantity grows with cooling solely via the jump length, $\Delta r(T)$, which also sets the amplitude of the long range elastic strain field. The full temperature dependence of the cooperative volume is rather complicated and can be obtained by substituting Eqs. (16) and (17) into the above relation. A simpler expression in the spirit of the present discussion is to use Eqs.(19) and (21), and ignore the weak logarithmic term in Eq.(20), to obtain

$$V_c \equiv F_{elastic}/G \propto N_s(B/T - A)^2 \qquad (22)$$

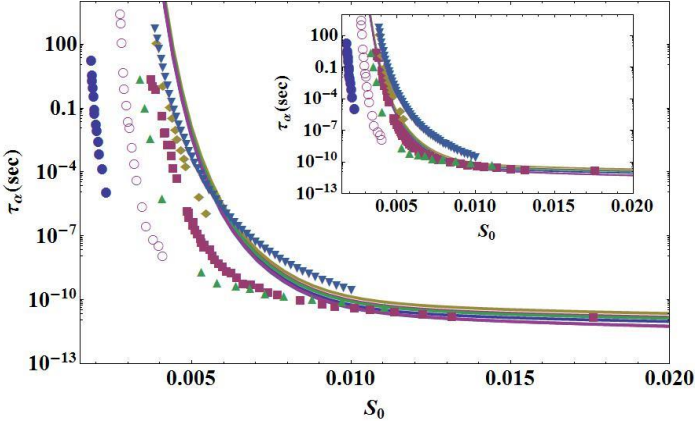

**Figure 2:** Alpha time as a function of dimensionless compressibility. Solid curves are theory results for toluene(blue), OTP(red), TNB(yellow), salol(green), glycerol(gray), and $S_0$ is the hard sphere compressibility. Experimental data is also shown for toluene(blue circles), OTP(red squares), TNB(yellow diamonds), salol(green triangles), glycerol(gray downward triangles), and sorbitol(open circles), where $S_0$ is the experimental molecular compressibility. Inset: Same plot but the theory curves have been shifted along the $S_0$ axis in order to better overlay the experimental data.

Within NLE theory, the localized state is associated with harmonic vibrations on the dynamic free energy. The corresponding frequency, which is a toy model for the Boson peak frequency, is [28]:

$$\omega_B = \frac{1}{2\pi}\sqrt{\frac{K_0}{M}}\left[1 - \frac{M}{K_0}\left(\frac{\zeta_s}{2M}\right)^2\right]^{1/2}$$
$$\approx \frac{1}{2\pi}\sqrt{\frac{K_0}{M}} \propto \sqrt{\frac{k_B T}{Md^2}}\left(\frac{d}{r_{loc}}\right) \qquad (23)$$

where $\zeta_s^{-1} = \beta Md^2\tau_s$, and the approximate equality has been established based on numerical calculations. This vibrational energy scale increases more slowly with cooling than G, though the localization length is the key quantity for both properties. The absolute magnitude of both G and $\omega_B$ depend on nonuniversal parameters. A caveat concerning Eq.(23) is the Einstein glass picture underlying NLE theory does not include a spectrum of phonon-like states, and one expects it over predicts (as we confirm below) the magnitude of the Boson frequency. However, interestingly, neutron experiments have found that the key features of the Boson frequency as deduced from incoherent (single particle) scattering and collective scattering are surprisingly similar [29, 30].

### B. Crossover Temperatures and Time Scales

Paper I discussed multiple theoretically well-defined characteristic or crossover volume fractions and their corresponding time scales. The initial crossover from the normal liquid to one where barriers are nonzero occurs corresponds to the (naïve) MCT transition at $\phi = \phi_A \approx 0.43$. From Eq.(12), this defines a temperature, $T_A$, where $S_0^{HS}(\phi_A) = 0.031$; the calculations in Table 1 show this temperature is far above $T_g$. However, the barrier initially grows in very slowly below $T_A$. A practical onset for activated dynamics is when the hopping time scale equals the renormalized binary collision time scale, thereby defining the crossover temperature $T_x$:

$$\tau_s(T_x) \equiv \tau_{hop}(T_x) \qquad (24)$$

Upon further cooling, the relative importance of the collective versus the local barrier grows. Two additional, theoretically well-defined crossover temperatures which indicate the change from a local hopping process to a collective hopping process are:

$$F_{elastic}(T^*) \equiv F_B(T^*) \qquad (25)$$
$$\frac{d}{dT}F_{elastic}(T') \equiv \frac{d}{dT}F_B(T') \qquad (26)$$

T' indicates where the growth rate of the collective barrier first exceeds that of the local barrier. As discussed below, it appears to correlate with diverse empirical estimates of the key dynamical crossover temperature. $T^*$ occurs at a lower temperature and its practical observable dynamic consequences will be shown to be much less pronounced.

### C. Dynamic Fragility

To gain intuition concerning what controls the dynamic fragility in our theory we perform an approximate analytic analysis which will be shown to accurately reproduce the key trends of our numerical calculations. The dynamic fragility is defined as

$$m \equiv \frac{1}{T_g}\frac{d\log(\tau_\alpha)}{d1/T}\bigg|_{T_g}$$
$$= \frac{1}{T_g}\frac{d\log(\tau_\alpha)}{d1/S_0^{HS}}\frac{1/S_0^{HS}}{1/T}\bigg|_{T_g} \qquad (27)$$

where the second expression uses the chain rule. From Eq. (14), one sees that $\log(\tau_\alpha) \approx F_{total}$ to a reasonable approximation at $T_g$ where $\tau_s \ll \tau_{hop}$. From Paper I it was shown that $F_{total} \approx F_B(1 + bF_B)$, prior NLE theory work numerically found $F_B \approx 0.08/S_0^{HS} - 3.51$, and from Eq. (8) one has $d(1/T)/d(1/S_0^{HS}) = 2N_s B(B/T - A)$. Employing all these results, one can write the "parabolic" relation: $\log(\tau_\alpha/\tau_s) = a_1 + a_2 S_0^{-1} + a_3 S_0^{-2}$ where $a_1$, $a_2$, and $a_3$ are to leading order constants. We note that $a_3 \propto (\Delta r_{eff})^2 \propto (\Delta r)^4$ quantifies the sensitive dependence of the collective barrier on the microscopic jump length that sets the amplitude of the strain field. Using all of the above results in Eq. (27) gives

$$m \propto \frac{1}{T_g}\left[a_2 + \frac{2a_3}{S_0^{HS}}\right] 2N_s B\left(\frac{B}{T} - A\right)\bigg|_{T_g} \quad (28)$$

$$\propto \frac{B}{T_g}\left[a_2 + \frac{2a_3}{S_g^{HS}}\right] 2N_s \left(\frac{B}{T_g} - A\right)$$

where $S_0^{HS}|_{T_g} \equiv S_g^{HS} \approx 0.0045$. Evaluating Eqs. (8) and (9) at $T_g$ gives $B/T_g = 1/\sqrt{N_s S_g^{HS}} + A$, and substituting this in Eq. (28) yields

$$m \propto \left(\frac{1}{\sqrt{N_s S_g^{HS}}}\right)\left[a_2 + \frac{2a_3}{S_g^{HS}}\right] \frac{2\sqrt{N_s}}{\sqrt{S_g^{HS}}}$$

$$\propto \left(\frac{2}{S_g^{HS}} + \frac{2A\sqrt{N_s}}{\sqrt{S_g^{HS}}}\right)\left[a_2 + \frac{2a_3}{S_g^{HS}}\right] \quad (29)$$

$$\propto 1 + cA\sqrt{N_s}$$

where $c$ is a numerical factor. The attraction strength parameter, B, sets the energy scale of $T_g$ but does not enter the fragility which is controlled to leading order by molecular size and entropic packing parameter A. Eq.(29) implies fragility is minimized when A<0 and the molecule is large (extended alcohols per Table 1), and is largest for big vdW molecules (A>0). However, it is the composite parameter, $A\sqrt{N_s}$, that controls the overall magnitude of the fragility within the present quasi-universal picture.

### D. Generic Low Temperature Limit and Strong Glass Forming Liquids

One can ask what the present theory predicts in a hypothetical T→0 limit. There is no Kauzman transition, but there is a jamming limit (random close packing at $\phi_J \approx 0.644$). For hard spheres, this corresponds to condensation into a sub-extensive number of inherent structures ("bottom" of the (free) energy landscape), and a crossover of the EOS from fluid-like to a free-volume-like form [31] where pressure diverges and dimensionless compressibility vanishes as [32, 33]: $\beta P/\rho \propto (\phi_J - \phi)^{-1}$ and $S_0 \propto (\phi_J - \phi)^2$. As T→0, a harmonic vibrational description should be generically relevant, though usually unattainable in equilibrium. However, some highly structured "strong" network glass formers (e.g., silica) that display Arrhenius relaxation [3] may effectively be in this low temperature regime with regards to their thermal density fluctuations even under equilibrated molten conditions.

Using the above scaling relations for P and $S_0$ in Eq.(10), the effective hard sphere volume fraction in the T→0 limit is

$$\phi \to \phi_J - c\sqrt{T} \quad (30)$$

where $c$ is a constant. Since density and isothermal compressibility approach limiting values as T→0, the dimensionless compressibility takes on a harmonic crystal form

$$S_0 \propto \rho k_B T \kappa_T \propto T \propto (\phi_J - \phi)^2 \quad (31)$$

We note that experiments [34] and simulations [35-37] on *molten* silica obey the linear scaling of $S_0$ with temperature in Eq.(31), a laboratory realization of "solid-like" behavior in the liquid phase. The form of Eq.(31) is not captured by typical fluid integral equation theories, and thus Eq.(16) does not apply. Rather, in the harmonic limit one must have

$$r_{loc}^2 \propto T \quad (32)$$

per neutron experiments at low temperatures [3, 30]. From Eq.(17) in the T→0 the jump length diverges as

$$\Delta r_{eff} \propto (r_B - r_{loc})^2 \propto |\ln(T)| \quad (33)$$

Using these results in Eq.(I.25) one obtains

$$\beta F_{elastic} \propto \frac{(\Delta r_{eff})^2}{r_{loc}^2} \propto \frac{(\ln(T))^2}{T} \quad (34)$$

Thus, a near Arrhenius behavior is *generically* predicted with logarithmic deviations that weaken the growth of the relaxation time relative to pure Arrhenius. This form seems to be qualitatively consistent with recent measurements [38] on equilibrated amber (a chemically complex but fragile liquid) below $T_g$. As a speculative comment, these results may also be relevant to the near Arrhenius behavior of very "strong" liquids as a consequence of their solid-state-like thermal dependence of the dimensionless compressibility in the liquid phase. We note that a near Arrhenius behavior below $T_g$ for fragile liquids is typically a nonequilibrium kinetic effect [3].

## IV. REPRESENTATIVE CALCULATIONS

We now numerically apply the theory to study the alpha time, glass transition temperature, dynamic fragility, shear modulus, "Boson-peak" frequency, and the characteristic crossover temperatures and times that can be *objectively* defined. By the latter we mean either they can be deduced unambiguously from the alpha time or from the 2-barrier theoretical picture. Comparisons with experiments are also presented.

### A. Barriers, Alpha Time and Characteristic Temperatures. General Aspects

Per section III, there are 4 theoretically well-defined crossover temperatures which are, in decreasing magnitude, $T_A$, $T_x$, T' and $T^*$. Fig. 3 illustrates these characteristic temperatures in the context of OTP. The main panel shows how the local barrier, collective elastic barrier, and total barrier distinctively, but smoothly, grow with cooling. The inset presents the ECNLE theory, local NLE only, and short time process relaxation times. The emergence of a barrier occurs at $T_A$ (literal NMCT transition) and corresponds to the high temperature beginning of the curve in the main figure. However, as shown in the inset, at this high temperature the barrier is so low that the timescale for activated hopping is *faster* than the renormalized binary collision timescale, $\tau_s$. Activated processes become important in a practical sense at $T_X$ (marked in the inset) when $\tau_s = \tau_{hop}$, which we find typically occurs when $\tau_\alpha \sim 10^{-10}$ s. Although the true activated process present above $T_X$ is subdominant, our theory nevertheless predicts an *apparent* Arrhenius behavior (barrier, $E_A$) due in part to the temperature dependence (e.g., thermal expansion) of $\tau_s$. This regime begins to break down at $T_{A,eff}$ (called $T^*$ by Kivelson and Tarjus [19, 39]) which we find is very close to $T_X$ where $\tau_\alpha \sim$ 10-100 ps (see Table 3) consistent with experimental data on many glass-formers [4, 5].

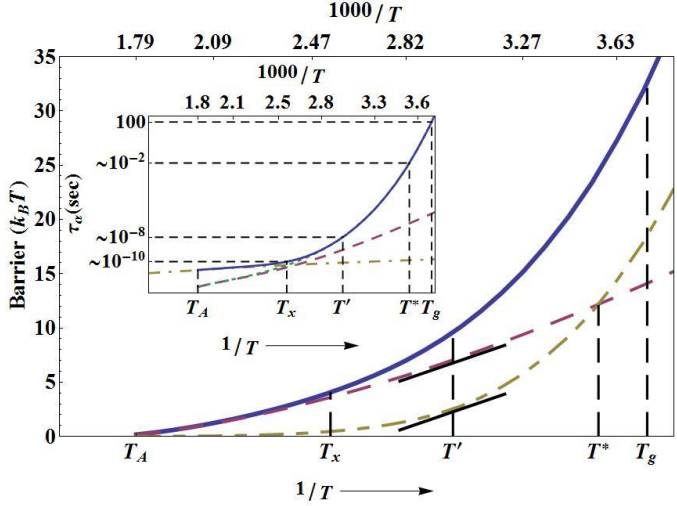

**Figure 3**: Schematic of theoretically important and well defined temperatures within the two barrier ECNLE picture. Although the plot is generic, the upper axis shows the absolute temperature scale for OTP as an example. The solid blue curve is the total barrier, the dashed red curve is the local barrier, and the lower dashed yellow curve is the elastic barrier, all plotted against inverse temperature. Inset: Alpha time as a function of inverse temperature with approximate timescales at characteristic temperatures marked. The solid blue curve is the alpha relaxation time of ECNLE theory, the dashed red curve is NLE theory analog, and the yellow dash-dot line is the dressed binary collision time..

Cooling below $T_X$, the next characteristic temperature is T' (see Eq.(26)) where the rate of thermal growth (temperature derivative of barriers) of the collective elastic barrier equals its local NLE analog. This temperature controls the practical observation of a rapid bending up of the relaxation time as a function of inverse temperature as the deeply supercooled regime is entered. In this regime $F_{elastic} \propto F_B^2$, and T' signals the crossover to when this quadratic relation applies. We suggest T' is the physical meaning of many empirically-extracted "crossover" or "onset" temperatures [1, 3] (e.g., $T_B$, $T_0$, $T_c$) identified in the literature which often occur when $\tau_\alpha \approx 10^{-7\pm1} s$. The final important fundamental temperature is $T^*$ where $F_{elastic} = F_B$. This temperature is quite low, and corresponds to $\tau_\alpha$ of order $10^{-2}$ s. As discussed further below, we believe this temperature is related to a reference temperature extracted by Rossler et al [22, 40], $T_R$, where the uncooperative barrier equals its collective analog. Finally, $T_g$ is identified with $\tau_\alpha = 100 s$.

### B. Experimental Comparisons

Quantitative application of our theory, using experimental EOS data to construct the mapping, has been performed for 5 nonpolar vdW molecules (toluene[41], biphenyl[42], OTP[43], TNB(assumed same as OTP), salol[44]), 3 alcohols (ethanol[45], glycerol[46], sorbitol[47]), 2 atomic metals (cesium[48], rubidium[48]) and two rare gases (argon[49], xenon[50]). Although the deeply supercooled regime and glass transition of the rare gases and liquid metals are not experimentally accessible, we present results for them as examples of different chemical classes, and also because there have been many simulations of such atomic systems. Numerical results for $T_g$, fragility and characteristic temperatures are shown in Table 1, temperature ratios (relative to $T_g$) in Table 2, and characteristic energy scales and time scales in Table 3. Where available, the corresponding experimental estimates are listed.

A comparison between a representative subset of our calculations of the alpha time as a function of temperature and $T_g$ values with experiments are shown in Fig. 4 (using experimental data for OTP[51], TNB[52], glycerol[53], salol[44], and toluene[54]) and Table 1, respectively. The computed $T_g$ values are generally within 20% of experiment [55-57], with toluene and sorbitol the biggest outliers; predictions for temperature ratios (Table 2) are more accurate than absolute values. In general, the temperature dependence of the alpha time seems remarkably accurate over 14 orders of magnitude in time from $\sim 2T_g$ to $T_g$ given the no adjustable parameter nature of the calculations. The largest deviation is for the hydrogen-bonding glycerol, as might be expected based on our use of a hard sphere model of structure and the short time process.

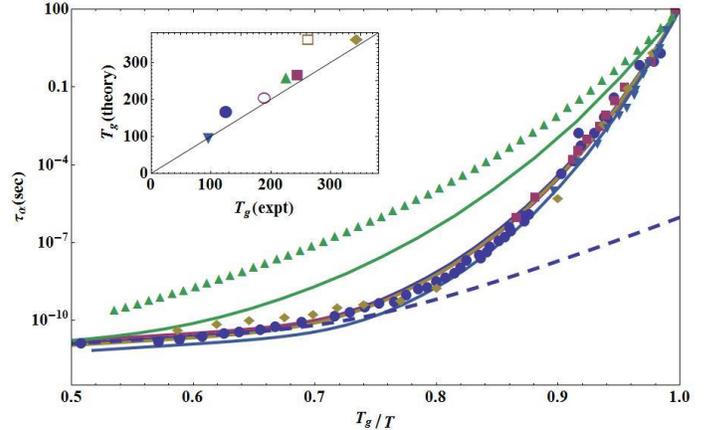

**Figure 4:** Angell plot showing theoretical calculations of the alpha time for OTP(blue), TNB(red), glycerol(green), salol(yellow), and toluene(gray). The theoretical result using only the local barrier is also shown for OTP(blue dashed line). Experimental data is shown for OTP(blue circles)[51], TNB(red squares)[52], glycerol(green upward triangles)[53], salol(yellow diamonds)[44], and toluene(gray downward triangles)[54]. Inset: Comparison of theory and experiment values of $T_g$. Points are from left to right: ethanol(down closed triangle), toluene(closed circle), glycerol(open circle), salol(upward closed triangle), OTP(closed square), sorbitol(open square), TNB(closed diamond).

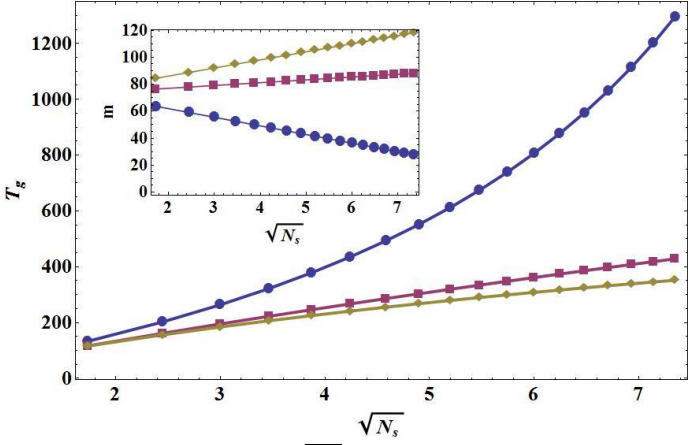

**Figure 5**: Theoretical $T_g$ vs $\sqrt{N_s}$ at fixed chemistry for glycerol(blue circles), OTP(red squares), toluene(yellow diamonds). Inset: fragility vs $\sqrt{N_s}$ at fixed chemistry. Symbols have the same meaning.

Table 1 shows a similar level of agreement with experiment applies to the dynamic fragility, with alcohols showing the poorest agreement, though not in a systematic direction. However, the theory clearly predicts a muted range of fragility compared to the real systems, perhaps again an expected consequence of using a hard sphere model for chemically diverse nonspherical molecules. For sorbitol, use of Eq.(2) instead of Eq.(8) shifts the fragility in the correct direction, but is still not quantitatively accurate. Nonetheless, our mapping to a reference hard sphere fluid does work reasonably well for glycerol which has fewer hydrogen bonds. Moreover, our results for the non-vdW molecule series of ethanol, glycerol, and sorbitol show the theory becomes quantitatively worse with increasing degree of hydrogen bonding, a sensible trend.

Fig. 4 also shows the relaxation time (dashed curve) for OTP using only the local NLE theory. In the supercooled regime, it appears almost straight, which in the inverse temperature representation implies apparent Arrhenius behavior over a wide temperature regime. However, such an apparent Arrhenius behavior is not experimentally observable in the deeply supercooled regime which is dominated by the collective barrier, and its existence cannot be rigorously deduced via extrapolation from the observed high temperature behavior which depends on both local hopping and binary collision physics. We note in passing that this NLE theory prediction in the deeply supercooled regime looks quite similar to an Arrhenius beta process(the Johari-Goldstein process), though we do not have a clear theoretical argument for the significance of this observation at present.

### C. Trends at Fixed Chemistry

We now perform model calculations at fixed chemistry, defined to be fixed values of A and B in Eq.(8), as relevant to a homologous series. The purpose is to cleanly expose the dependence of $T_g$ and fragility on molecular size as encoded in the number of rigidly moving sites, $N_s$. The main frame of Fig. 5 shows that roughly $T_g \propto \sqrt{N_s}$ for vdW molecules per the analytic analysis in section III and experiment [25], but grows much faster for glycerol. This reflects the opposite sign of A in the dimensionless compressibility in Eq.(8). Also, per the analytic analysis (see Eq.(29)), the inset to Fig. 5 shows the fragility is linearly related to $A\sqrt{N_s}$. The opposite (sign) change with increasing $N_s$ reflects the thermodynamic difference of the EOS of glycerol and vdW molecules which enters via the opposite signs of the entropic parameter A.

### D. Short Time Properties

Fig. 6 presents representative calculations of two short time properties, the elastic plateau shear modulus and Boson peak frequency. Results are shown in absolute units as a function of scaled temperature for one vdW and one hydrogen-bonding molecule. These properties probe the localization well region of the dynamic free energy, and are quantitatively sensitive to nonuniversal factors such as molecule diameter and mass.

The magnitude of the shear modulus of glycerol is within a factor of 3 of experiment [58], and its increase with cooling is rather well predicted. Similarly, the Boson peak frequency is also over predicted (by a factor of 3 or 5), though the weak temperature dependence is again reasonably well captured. Given in our theory the barriers are tightly related to the localization length, the temperature dependence of these results are important for our description of super-Arrhenius relaxation.

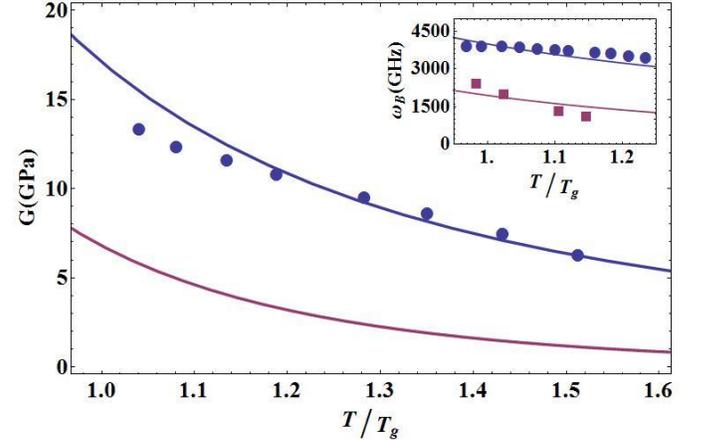

**Figure 6:** Shear modulus in GPa as a function of temperature for glycerol(blue) and OTP(red) using a molecular diameter (d) of 6.14 Å and 9.1 Å for glycerol and OTP, respectively. Closed blue circles are glycerol experimental data [58], shifted upward by a factor of 2.75. Inset: Boson peak frequencies for glycerol(blue), OTP(red). Experimental data for glycerol(blue circles, shifted upward by 3)[14] and OTP(red squares, shifted upward by 5).

## V. CONNECTION TO ELASTIC MODELS

Aspects of ECNLE theory have clear similarities to Dyre's phenomenological elastic shoving model [11, 12, 59] which postulates that super-Arrhenius behavior is due solely to a (plateau) shear modulus that grows with cooling:

$$\tau_\alpha \sim \tau_0 e^{\frac{G v_c}{k_B T}} \quad (35)$$

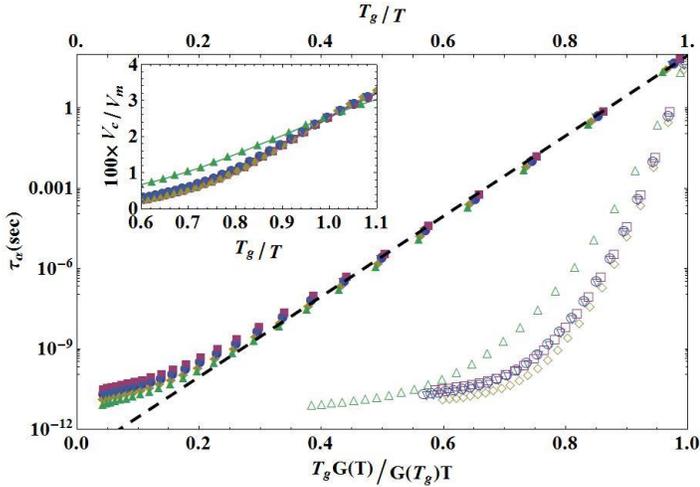

**Figure 7:** Alpha time calculations for OTP(blue circles), TNB(red squares), toluene(yellow diamonds), glycerol(green triangles), and salol(gray downward triangles). Closed symbols are plotted versus the dimensionless variable X (lower axis) defined in the text and as suggested by the shoving model [11,12], open symbols are the same calculations plotted versus $T_g/T$ (upper axis). Inset: Ratio of cooperative volume to molecular volume plotted against $T_g/T$. Symbols have the same meaning.

where the "cooperative volume", $v_c$, is a material-specific, temperature-independent fit parameter. If this formula is consistent with our theory, then plotting our calculations against the normalized quantity $X \equiv T_g G(T)/TG(T_g)$ should lead to a universal collapse of different systems onto a single line. Fig. 7 shows typical results of such a comparison. Excellent agreement is found over the slowest ~10 orders of magnitude of relaxation, despite the fact ECNLE theory has a growing correlation volume with cooling (see inset) and a local non-cooperative barrier. The upward deviation at higher temperatures (corresponding to a relaxation time ~$10^{-8}$ sec) is consistent with experimental data analysis [11,12,58], and finds a precise interpretation within ECNLE theory as due to the dominance of the local barrier and binary collisions at high temperatures. The inset to Fig. 7 shows, consistent with phenomenological estimates [11,12,58], the theoretically computed cooperative volume, $V_c$, is smaller than molecular size, but grows by a factor of ~2-3 over the range of temperatures studied for all materials. As discussed in paper I, this growth with cooling addresses a key criticism [2] of the elastic shoving model.

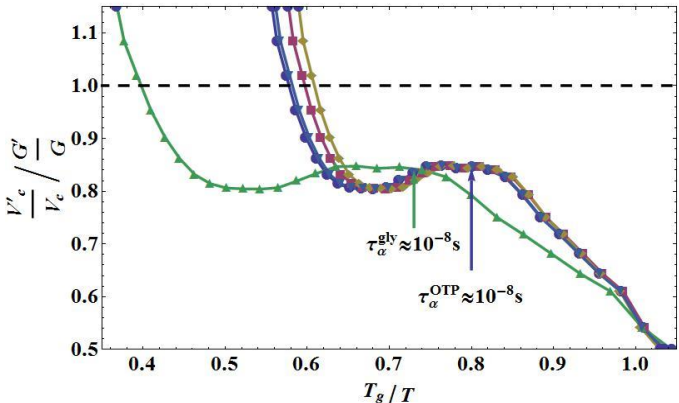

**Figure 8:** Ratio of logarithmic derivatives of the two contributions to collective elastic barrier versus $T_g/T$ for OTP(blue circles), TNB(red squares), toluene(yellow diamonds), glycerol(green triangles), and salol(downward gray triangles).

It is of interest to ask what is the dominant contribution to the growth in the collective elastic barrier of ECNLE theory over the regime in which it controls the alpha relaxation time? This can be determined based on logarithmic derivatives

$$\frac{\partial F_{elastic}}{\partial T} = GV_c\left(\frac{\partial \ln G}{\partial T} + \frac{\partial \ln V_c}{\partial T}\right) \quad (36)$$

Representative results for glycerol and OTP are shown in Fig. 8, where a prime indicates a temperature derivative; the temperature at which the relaxation time reaches $10^{-8}$ s is marked by an arrow. At higher temperatures, reference to Fig. 3 shows that the growth of the elastic barrier is subdominant and (see Fig. 7) the phenomenological shoving model breaks down. By comparing the ratio of the two terms inside the parentheses of Eq. 36 one sees that in the deeply supercooled regime the dominant contribution to collective barrier growth comes from the change in the shear modulus. This helps explains why the phenomenological shoving model can work well in the supercooled regime. At higher temperatures competing factors become important and the shoving model breaks down.

## VI. PRESSURE EFFECTS and GROWING COOPERATIVE LENGTH SCALE

We now consider how pressure modifies the temperature dependence of the alpha time, $T_g$, and fragility based on using Eq. (2) directly as $S_0^{HS}(\phi_{eff}) \equiv S_0^{expt}(T;P)$ and EOS data. Knowledge of the pressure dependence can be employed to also compute an activation volume that appears to track dynamical cooperativity.

**A. Pressure Dependence of the Alpha Process**

Fig. 9 shows a representative result for the variation with pressure of the alpha time of OTP. With increasing pressure, relaxation in the supercooled regime is slowed far more than at high temperatures. The inset shows fragility and $T_g$ results for three vdW liquids and glycerol. Except for toluene, all systems exhibit a monotonic and slightly sub-linear growth $T_g$ with pressure. The unusual behavior of toluene occurs only at very low pressures, and we are unsure whether the employed EOS input is reliable. All systems show a decrease of fragility with pressure, which agrees with experiment [60] except for glycerol where fragility increases with pressure.

Quantitatively, the theory generally predicts a stronger sensitivity to pressure than observed. This is perhaps unsurprising given the "free-volume-like" nature of mapping thermal liquids onto an effective hard sphere fluid. Moreover, the incorrect sign of the fragility dependence for hydrogen-bonding glycerol is again likely not unexpected. Quantitatively, the change in $T_g$ with pressure is overestimated by about a factor of ~2 for the fragile vdW liquids and ~4 for glycerol. The quantitative sensitivity can be described by the value of the derivative $dT_g/dP$. The calculated (measured) values in units of K/MPa are: ~0.47(0.26[61]) for OTP, ~0.1(unknown) for toluene, 0.16(0.04[60]) for glycerol, and 0.4(0.2[60]) for salol. The predicted (measured) fragility pressure derivatives as P→0 (in inverse MPa) are: -0.24 (-0.24[60]), -1.2 (unknown), -0.038 (0.035[60]), and -0.2 (-0.11[60]) for OTP, toluene, glycerol and salol, respectively.

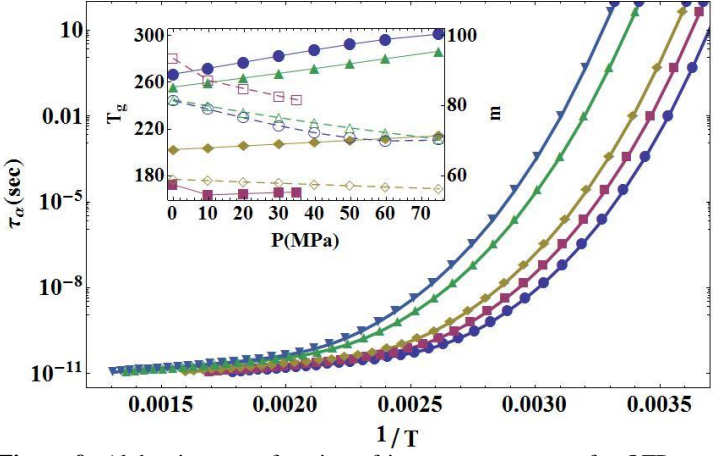

**Figure 9:** Alpha time as a function of inverse temperature for OTP at pressures (from left to right) of 0.101, 10, 20, 50, 75 MPa. Inset: $T_g$(right axis) and fragility(left axis) vs pressure for OTP(blue circles), toluene(red squares), glycerol(yellow diamonds), and salol(green triangles). Closed symbols are $T_g$(left) and open symbols are fragility(right)

### B. Activation Volume and Growing Length Scale

Sokolov and coworkers [14] have recently studied the pressure and density dependence of the structural relaxation time based on the concept of a temperature-dependent activation volume, $\Delta V^{\#}$, defined as

$$\tau_\alpha(T,P) \equiv \tau_\alpha(T,0)\exp\left(P\Delta V^{\#}(T)/k_B T\right) \quad (37)$$

The physical idea is pressure enters in a mechanical work fashion quantified by a temperature-dependent "activation volume" that reflects the degree of molecular level re-arrangement required for the alpha process. In practice, the latter is computed as

$$\Delta V^{\#}(T) \equiv k_B T \frac{d}{dP}\ln\left[\tau_\alpha(T,P)\right]\bigg|_{P\to 0} \quad (38)$$

and thus is fundamentally a response-like quantity. Experimentally, the activation volume at $T_g$ was found to be [14] ~ 0.67, 0.44 and 0.06 $nm^3$ for OTP, salol and glycerol, respectively; this corresponds, e.g., to ~4 times the OTP molecular volume. Our corresponding theoretical calculations using Eq.(38) are 1.28, 0.96, and 0.28 $nm^3$. Relative trends are well predicted, and the level of quantitative disagreement is very similar to our calculations for $dT_g/dP$, consistent with the exact relation: $V^{\#}(T_g) = mR(dT_g/dP)/\log(e)$.

Results for the full temperature dependence of $\Delta V^{\#}(T)$ are shown in Fig. 10 in two formats for OTP, salol and glycerol. The inset presents the absolute value of the activation volume versus reduced temperature. After modest vertical shifting (by factor of ~0.55, 0.5 or 0.25, corresponding roughly to our overestimate of the change in $T_g$ with pressure), one sees theory and experiment are in excellent agreement. The main frame of Fig.10 explores the possibility of a near universal collapse as a logarithmic function of the alpha time. The theory results collapse essentially perfectly, as expected based on the effective hard sphere fluid mapping. The experimental data also collapse well, and the slope of the logarithmic dependence is very close to what is predicted. Differences between theory and experiment are seen at high temperatures outside the deeply supercooled regime where the alpha time is less than ~ $10^{-8}$ s. We believe the above results provide strong support for both the ECNLE dynamical ideas and our mapping.

Very interestingly, the activation volume appears to be tightly correlated with measures of dynamic heterogeneity and correlation. Specifically, the cube root of the activation volume at $T_g$ has been shown to correlate with the characteristic length scale of the Boson peak [13, 14]. In addition, $\Delta V^{\#}(T)$ appears to have the same temperature dependence as the number of correlated molecules, $N_{corr}(T)$, as extracted from nonlinear dielectric measurements for glycerol [62]. Both these quantities increase as a logarithmic function of $\tau_\alpha$ in the deeply supercooled regime, by roughly a factor of 2 over a temperature range where $\tau_\alpha$ grows by 9 decades (from ~$10^{-8}$ to 10 s). Precisely the same behavior of $\Delta V^{\#}(T)$, including a universal collapse, has been found for OTP, salol and polystyrene [14]. Collectively, these studies support the suggestion of a generic connection between a heterogeneity or cooperativity volume and the activation volume.

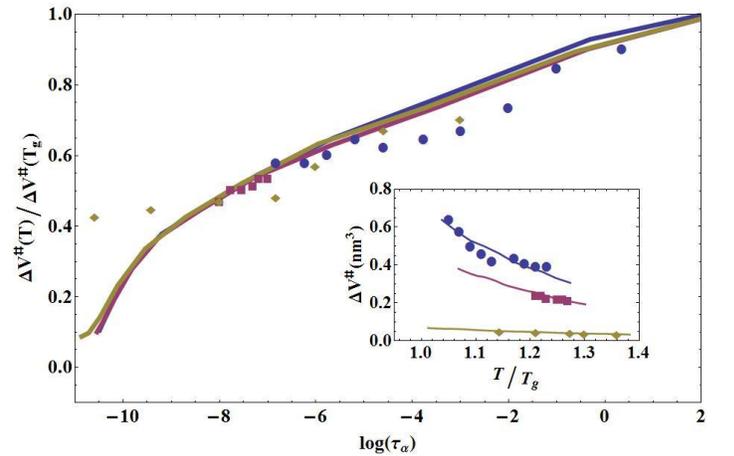

**Figure 10:** Activation volume of Eq. (38) normalized to unity at $T_g$, plotted against the logarithm of the alpha time for OTP(blue), salol(red), and glycerol(yellow). Experimental data [14] are shown for OTP(blue circles), salol(red squares), and glycerol(yellow diamonds). Inset: Activation volume in units of $nm^3$ plotted against $T/T_g$ with symbols retaining the same meaning. The theoretical calculation for OTP has been scaled by a factor of 0.5, for salol by a factor of 0.55, and by glycerol by a factor of 0.25 as discussed in the text.

Most recently, the nonlinear dielectric measurement of $N_{corr}(T)$ for 4 liquids of widely variable chemistry and fragility have been reported [63]. In all cases it was found

$$\tau_\alpha(T) \propto \exp(bN_{corr}(T)) \quad (39)$$

where $b$ is a system-specific numerical factor. Given Eq.(39), and that prior experiments [13] generically found $\Delta V^\#(T) \propto \ln\tau_\alpha(T)$ in the deeply supercooled regime consistent with our theory, an effective barrier proportional the "number of correlated molecules" can be viewed as a logical inference of the ECNLE approach. However, the notion of a number of correlated particles determining the barrier, a central concept of models based on compact domains of re-arranging particles, does not directly enter ECNLE theory.

As discussed in paper I, there are other growing length scales in ECNLE theory, albeit not directly experimentally measurable and/or model-dependent. For example, $V_c(T)$ in the inset of Fig.7, or the microscopic jump length which sets the amplitude of the long range elastic strain field. All these quantities grow slowly with cooling (more or less logarithmically) by modest factors in the supercooled regime, though they differ physically and with regards to the quantitative temperature sensitivity.

## VII. CONNECTIONS TO ARRHENIUS, MODE COUPLING, ENTROPY CRISIS, DYNAMIC FACILITATION AND OTHER MODELS

We now treat our theoretical calculations as "data" and analyze them in the context of diverse models (as done by experimentalists and simulators). Our goal is to see how our predicted temperature dependence of the alpha time in various "regimes" compares with different models, and extract empirical characteristic temperatures.

### A. Models

We consider four classes of models that aim to describe: (1) a high temperature (apparent) Arrhenius regime, (2) a narrow intermediate crossover regime, (3) the deeply supercooled regime, and (4) all regimes. At high temperatures and fast relaxation times, an apparent Arrhenius law is often found to fit experiments [4,19, 22,39,40,64]:

$$\tau_\alpha \propto \exp(E_A/k_BT) \quad (40)$$

where $E_A$ is often many times (~ 5-6 for molecules) the thermal energy, strongly suggesting it is not solely a thermal expansion effect. Ideal MCT [15] has been proposed to describe a narrow intermediate crossover window between the high temperature and deeply supercooled regimes over 3 or so decades where $\tau_\alpha \approx 10^{-10} - 10^{-7} s$. The alpha time is a critical power law, with a hypothetical (unphysical) divergence $T_c$:

$$\tau_\alpha \propto (T - T_c)^\nu \quad (41)$$

Two distinct thermodynamic entropy crisis approaches for the deeply supercooled regime are the Adams-Gibbs (AG) model [16] and Random First Order Theory (RFOT) [17]. The former builds on a high temperature local activated event as the basic excitation, while the latter does not. In both cases, configurational entropy controls the barrier in the deeply supercooled regime leading to the classic VFT form (also motivated from very different "free volume" arguments [65]):

$$\tau_\alpha = \tau_0 \exp\left[\frac{D}{T - T_{vft}}\right] \quad (42)$$

where $T_{vft} = T_K$ in the literal Kauzmann paradox (zero configurational entropy) scenario. The VFT formula has three adjustable parameters, and fails at high enough temperature. Some have suggested [66-68] a two VFT ad hoc model corresponding to different high and low temperature VFT fits with a crossover at $T_B$. "Regimes" are identified based on the "Stickel analysis" [67] where $[d\log\tau_\alpha/d(T_g/T)]^{-1/2}$ is plotted against $T^{-1}$. In this representation, a Arrhenius law is a horizontal line and the VFT law is a straight line; the intersection of the high and low temperature versions of the latter defines $T_B$ (often close to [69] the empirically-extracted MCT $T_c$), while extrapolation of the low temperature form to zero empirically defines a hypothetical zero mobility state at $T_{vft}$.

Coarse-grained dynamic facilitation models based on directional mobility field propagation predict a "parabolic law" in the deeply supercooled regime [18, 70, 71]:

$$\log(\tau_\alpha/\tau_0) = \frac{J}{kT}\left(\frac{T_o}{T} - 1\right)^2, \quad T < T_o \quad (43)$$

where the mobile defect creation energy, $J$, and onset temperature, $T_0$, are determined by data fitting [70]. There are no divergences above T=0, and in the low temperature limit Arrhenius behavior emerges as the defect concentration approaches zero.

Tarjus and Kivelson [19, 20, 39], Rossler and coworkers[22, 40] and others[72] have suggested phenomenological 2-barrier models where the high temperature process is Arrhenius. Rossler et.al. have convincingly shown this picture can empirically fit relaxation data on many molecular liquids over 14 orders of magnitude based on [22, 40]:

$$\tau_\alpha = \tau_\infty \exp\left(\frac{E_A + E_{coop}(T)}{k_BT}\right)$$
$$\approx \tau_\infty \exp\left(\frac{E_A}{k_BT}\left[1 + e^{-\lambda(\frac{k_BT}{0.1E_A}-1)}\right]\right) \quad (44)$$

where the λ is nonuniversal parameter that is weakly varying for vdW liquids (e.g., $\lambda \approx 7.8 \pm 0.7$ for toluene, salol, OTP, TNB). Detailed data analysis based on global fits of 18 liquids over 14 orders of magnitude in alpha time suggests remarkable connections between the Arrhenius and cooperative barriers. These deductions are relevant to testing our theory, and the key trends are as follows [22, 40]. (a) A crossover temperature, $T_R$, is defined as when the uncooperative local and cooperative barriers are equal: $E_{coop}(T=T_R) \equiv E_A$; a near universality is found for 15 vdW liquids, $T_R/T_g \approx 1.13 \pm 0.02$, with larger values found for less fragile alcohols (e.g., $T_R/T_g \approx 1.26$ for glycerol). (b) Near universal values are found for the two barriers relative to the glass transition temperature: $E_A/k_B T_g \approx 11 \pm 1$ and $E_{coop}(T_g)/k_B T_g \approx 24 \pm 1$, and thus at $T_g$: $E_{coop}/E_A \approx 2.2$. Interestingly, this implies the system needs to be quite close to kinetic vitrification before the collective barrier exceeds its apparent Arrhenius analog, and even at $T_g$ the former is only about twice as large as the high temperature barrier. (c) The "prefactor" in Eq.(44) is physically sensible, $\sim \tau_\infty \approx 10^{-13 \pm 0.3} s$, of order 0.1 ps. (d) Though not experimentally measurable, Eq.(44) predicts a finite low temperature cooperative barrier corresponding to a return to Arrhenius behavior $E_{coop}(T \to 0) = E_\infty \left(1 + e^{\mu b}\right)$, where $\mu b \approx 3-8$.

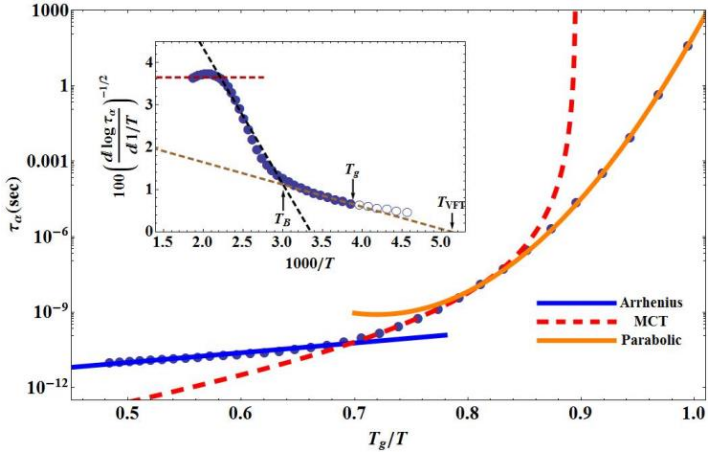

**Figure 11:** Analysis of the theoretical alpha time for salol(blue circles) in different regions in terms of a high temperature Arrhenius law(blue), a crossover MCT region(red dashed) and a low temperature parabolic law(orange). Inset: Corresponding Stickel plot that shows a high temperature Arrhenius regime and two VFT laws with intersection at $T_B$. The open circles represent the theoretical calculation below $T_g$, showing that the extrapolation to $T_{VFT}$ is only an apparent feature of the theory.

### B. General Findings

We now treat our theoretical calculations as "data" and fit them to above forms. A representative result is shown in Fig. 11 for salol. One can indeed interpret our calculations in the commonly adopted "3 regime" scenario: a high temperature Arrhenius regime, a narrow crossover regime described by a critical power law, and a deeply supercooled regime where, e.g., the parabolic law fits our calculations very well. Of course, such a three regime picture is not literally part of our approach where the alpha time over 14 orders of magnitude comes from a single physical theory.

Based on the Stickel analysis, the inset to Fig. 11 shows that the theoretical data can also be well fit by a high temperature Arrhenius law, and two VFT laws with a crossover at $T_B$, per experimental analyzes [67, 68]. By means of an ad hoc extrapolation, an apparent divergence of $\tau_\alpha$ at $T_{VFT}$ can be extracted from the low temperature regime. In reality, there is no finite temperature divergence in our theory, and our calculations upwardly deviate from the VFT law just below $T_g$.

Characteristic temperatures, and their ratios compared to $T_g$, associated with the various fits to our theoretical calculations are listed in Tables 1 and 2; the numbers are reasonable with regards to their experimentally-extracted analogs. We caution that extracting characteristic times and temperatures via fitting introduces an element of subjectivity as to what constitutes a "good" fit. The Tables show that the theoretically well-defined temperature $T_X$ is associated with the empirically-deduced end of Arrhenius behavior at $T_{A,eff}$. The theoretical T' defined in Eq. 26 is associated with the important and physically meaningful crossover to cooperative dynamics, here precisely defined as when the temperature growth of the net barrier begins to be controlled by collective elasticity; we suggest T' is the physical meaning of the empirical $T_B$. One can associate the parabolic law $T_0$ with either $T_X$ (since one can extend a parabolic law fit down to $T_{A,eff}$ though some fit quality is sacrificed) or T'. Conceptually, it seems best to associate it with T', but we find that the "best" empirical fit lies somewhere between these temperatures.

Because of the underlying universality of our mapping to a hard sphere fluid, the time scales associated with the characteristic temperatures discussed above are only weakly material dependent (variations of typically one order of magnitude due to the system-specific short timescale); examples are given in Table 3. The end of the apparent Arrhenius regime is experimentally reported [4, 22, 40] to lie at $\tau_\alpha \approx 10^{-10.8 \pm 0.5} s$, and typically describes only 1 decade or less of the alpha time growth, features in good accord with our results. Upon further cooling an important dynamical crossover occurs at $T_B$, $T_c$, or $T_0$ where $\tau_\alpha \sim 10^{-8}$-$10^{-6}$ s.

In this section we have focused on the big picture and summarized the highlights of our comparisons. We now consider each of the regimes in more detail.

### C. Apparent Arrhenius and Intermediate Regimes

Fig. 11 shows an apparent Arrhenius law is predicted over a wide high temperature window, and begins to "fail" at $T_{A,eff} \sim 1.4\, T_g$ where $\tau_\alpha \approx 10^{-10} s$. A narrow, roughly 3 decades in intermediate time regime can then be fit using a MCT critical power law. However, its physical significance is unconvincing for at least two reasons. First, we know our "data" reflects activated hopping! Second, the non-singular parabolic law fits the slowest 12 orders of magnitude very well before failing at $\tau_\alpha \approx 10^{-9} - 10^{-8} s$.

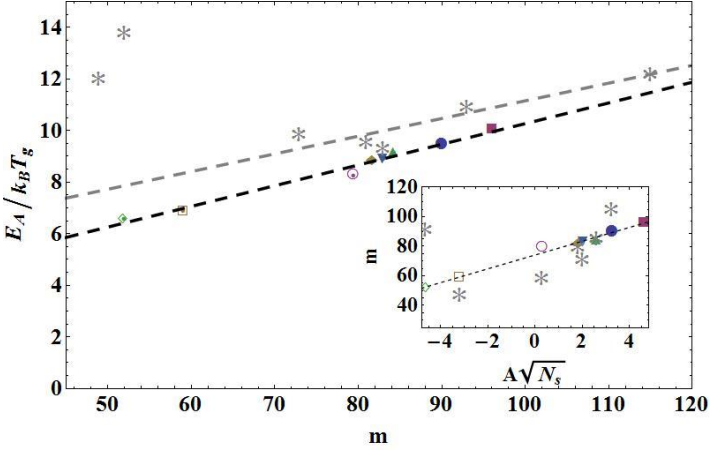

**Figure 12:** Apparent Arrhenius barrier plotted versus fragility. Colored points are the theoretical calculations for, from left to right: sorbitol, glycerol, ethanol, OTP, salol, TNB, toluene, and biphenyl. Gray stars are experimental data from left to right: glycerol[40, 56], propylene glycol(PG)[40, 56], salol[40, 56], OTP[40, 56], TNB[40, 56], propylene carbonate(PC)[40, 56], and toluene[40, 55]. The two outliers are strongly hydrogen bonding. Inset: Fragility plotted against the theoretical control variable. Colored symbols are the theoretical calculation, gray stars are experimental data. Within each set, the points are, from right to left, sorbitol, glycerol, ethanol, OTP, salol, TNB, toluene, and biphenyl.

Table 3 presents our extracted apparent Arrhenius barriers, in absolute units and relative to the glass transition and effective crossover temperatures. Recall the apparent Arrhenius behavior does not correspond to a pure barrier hopping process, but rather reflects the combined consequences of an effective binary collision process and low true barriers. In the absence of activated processes, we find the apparent barrier due to only dressed binary collisions is ~$5k_BT_g$, while the full calculation gives an apparent barrier of ~8-9 $k_BT_g$. The ratio of the apparent barrier to the effective crossover temperature ranges from 4-7, emphasizing that this apparent barrier energy scale is not small compared to the temperature interval over which an apparent Arrhenius behavior is extracted.

The main frame of Fig. 12 plots $E_A / k_B T_g$ versus fragility and shows we predict $m \propto E_A / k_B T_g$. This is also found for the shown experimental data [40, 55, 56] except for the two strongly hydrogen bonding systems, for which our fragility results are not accurate. The inset of Fig.12 plots fragility versus the quantity our theory predicts controls it. The plot demonstrates Eq.(29) describes the numerical ECNLE theory results very well, although the ability of it to correlate the experimental data does not appear as strong due to the muted range of fragilities we predict based on our hard sphere mapping.

**D. Entropy Crisis Perspective**

Entropy crisis [16, 17] and free volume [65] theories assert the alpha time diverges at a nonzero temperature. Based on the Stickel analysis, our calculations of the dynamical divergence temperature obtained by fitting and extrapolation are listed in Tables 1 and 2. The predicted ratios of $T_{vft}/T_g$ ~ 0.66-0.8 are in a range consistent with the rough experimental estimates [71, 73, 74] for these materials.

Adams and Gibbs argued the alpha time is a magnified version of an underlying Arrhenius (single particle or un-cooperative local) barrier hopping process per Eq.(I.48). The effective barrier $E_{eff} = z(T)E_A$, where $z(T)$ describes an increasing number of particles that participate in the alpha event upon cooling which, based on the presumed existence of a thermodynamic Kauzmann transition, leads to $z(T_K) \to \infty$. However, experiments suggest only modest values of z ~ 3-5 even at $T_g$.

We analyze our theoretical data in the AG-like spirit in two different ways. First, per Eq.(I.50), a well-defined theoretical approach based on our two barrier picture is :

$$z(T) = 1 + \frac{F_{elastic}(T)}{F_B(T)} \qquad (45)$$

Alternatively, since barriers are not observable, a pragmatic approach often employed in simulation and experimental studies is to identify the degree of effective cooperativity as

$$z(T) = W \frac{\ln(\tau_\alpha / \tau_s)}{E_A} \qquad (46)$$

where the numerical factor W is chosen such that z→1 at high temperature, consistent with the empirical extraction of an apparent Arrhenius barrier.

The main frame of Fig. 13 shows ECNLE theory calculations of z(T) based on Eqs.(45) and (46) for two representative systems. With cooling, one sees a smooth growth with z→ ~2.4 or ~3.5-5 at $T_g$ depending on which metric is used. The inset plots the inverse cooperativity parameter versus temperature down to $T_g$. Although z(T) never diverges (one can see the curvature below $T_g$ in the plots), in the Adam-Gibbs entropy crisis spirit we linearly extrapolate its inverse to zero to estimate a dynamic analogue of $T_K$. Results obtained from both approaches are given in Table 1; the numbers are reasonable, and bracket the VFT extrapolated dynamic divergence temperature. Table 2 shows the ratio of the *mean* $T_K$ to $T_g$ agrees quite well with (often imprecise) experimental estimates. The deduced ratios of $T_K/T_{vft}$ also seem reasonable; recall that in experiment they are sometimes close to unity, but are known to show significant deviations in both directions for diverse materials [71, 74]. We emphasize that in our approach there are no true divergences, so we ascribe no physical significance to the extrapolations. However, the sensibility of our extracted divergence temperatures compared to experimental estimates is meaningful.

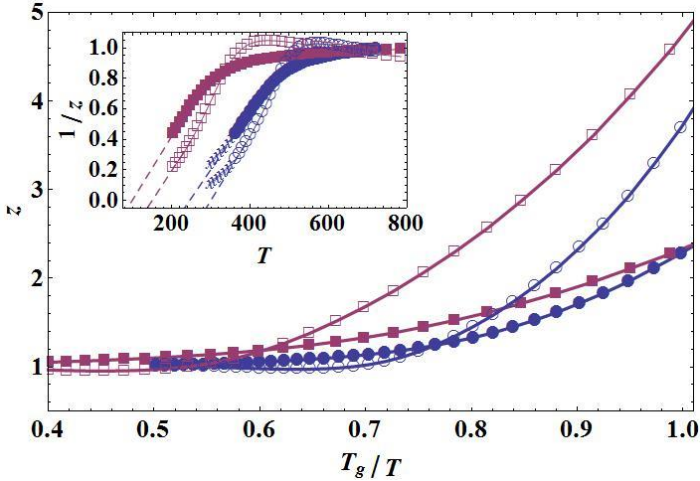

**Figure 13:** Theoretical cooperativity parameter, z, as a function of $T_g/T$ for glycerol(blue circles) and TNB(red squares); the TNB results are representative of what we find for all vdW liquids studied. The closed symbols are calculations based on Eq. (45) and the open symbols employ Eq. (46). Inset: Inverse cooperativity parameter versus temperature and its linear extrapolation to zero to extract an apparent Kauzman temperature. The symbols retain the same meaning. The extrapolation using Eq. (45) is labeled $T_K(alt)$, while Eq. (46) is employed to extract $T_K$. The "x" symbols are the cooperativity parameter for TNB below $T_g$ showing that the extrapolation of $z^{-1} \to 0$ is not really justified.

### E. Dynamic Facilitation

Values of the extracted defect energy (J) and onset temperature ($T_o$) based on parabolic law fits are shown in Tables 1 and 2; Table 3 lists J in units of $k_B T_o$, and the alpha time at $T_o$. All the extracted numbers seem very reasonable compared to prior fits of Eq.(43) to experimental data [70]. Overall, Eq.(43) provides a remarkably good and consistent fit of our theoretical calculations in the deeply supercooled regime. However, the physics underlying ECNLE theory is not dynamic facilitation, at least not in the sense of a literal conserved population of mobile defects.

We emphasize that in the ECNLE framework the "parabolic law" idea is not unique. Rather, there are multiple versions in the sense that the total barrier can be expressed as a quadratic function of diverse control variables (both dynamic and static), which all accurately capture our numerical results in the deeply supercooled regime. These control variables include: (i) the compressibility factor $Z = \beta P/\rho$ (Eq(I.41)), (ii) the local barrier $F_B$, (iii) the inverse dimensionless compressibility (or bulk modulus) $S_0^{-1}$, (iv) the inverse temperature, and (v) the inverse localization length $r_{loc}^{-1}$. Within ECNLE, the fundamental variable is the dynamic $r_{loc}$, and all other representations are consequences of the degeneracy between these variables, established theoretically via the ultra-local limit analysis discussed in paper I. Because we take the hard sphere fluid to be a quasi-universal model, these relations are carried over directly to thermal liquids via our mapping.

### F. Two-Barrier Phenomenological Models

Underlying Eq.(44) is a crossover temperature, $T_R$, where the apparent Arrhenius and cooperative barriers are equal. In terms of ECNLE theory, this corresponds to: $E_A = E_{coop}(T_R) \equiv F_B(T_R) + F_{elastic}(T_R) - E_A$. Tables 1 and 2 show our calculations of $T_R$. The values agree well with experimental results for 15 vdW molecules that found [22, 40] the nearly universal result $T_R/T_g \approx 1.13 \pm 0.02$; significantly larger values are observed (and predicted) for alcohols. Table 3 shows the alpha time at $T_R$ can vary by roughly 4 orders of magnitude. One also sees from Table 3 the predicted apparent high temperature Arrhenius barriers agree quite well with those extracted experimentally, including the nearly universal value of $E_A/k_B T_g \sim 11$ for vdW molecules [22, 39, 40]. The experimental estimate of $E_{coop}/E_A \approx 2.2$ at $T_g$ suggests an AG parameter of $z \sim 3.2$, consistent with our calculations.

Overall, we conclude that the ECNLE theory form of $\tau_\alpha(T)$, and the characteristic temperatures, energy scales and time scales extracted from it, are consistent with diverse glassy dynamics models. Given these diverse models generally claim good agreement with experiment based on multi-parameter *fits*, we feel this provides support for the accuracy of our approach. However, we emphasize that our physical picture is fundamentally different than the models discussed in this section, involves no fitting parameters, and is applicable in all "regimes".

### VIII. SUMMARY AND DISCUSSION

We have proposed a mapping from thermal liquids to an effective hard sphere fluid based on matching the thermodynamic-state-dependent dimensionless "long" wavelength amplitude of density fluctuations, or compressibility. Coupled with the ECNLE theory of the alpha relaxation in hard sphere fluids, this mapping results in a zero adjustable parameter theory that can be applied to quantitatively treat alpha relaxation over 14 or more decades in time in a unified manner. The theory has no singularities above zero Kelvin, and relaxation in the equilibrium low temperature limit is predicted to be of a roughly Arrhenius form due to condensation of the liquid into the bottom of the potential energy landscape.

The basic excitation in the theory is of mixed local-nonlocal spatial form reflecting a cage scale activated process which requires a long range collective elastic fluctuation in order to occur. This leads to a two-barrier description that is the key to the rich dynamic behavior predicted, encompassing apparent Arrhenius, crossover, and deeply supercooled "regimes", and multiple time and temperature characteristic crossovers. The theory also has multiple growing length scales upon cooling which reflect distinct aspects of the activation event. Of special interest is the experimentally measurable activation volume [13, 60], which is accurately predicted and correlates with a dynamic heterogeneity length scale and also the number of correlated particles as deduced from nonlinear dielectric and other measurements[62,63].

The calculated effects of pressure are qualitatively reasonable, although systematically too large. The theory also predicts a narrower range of dynamic fragilities than observed in thermal liquids. We believe these quantitative inaccuracies are likely unavoidable given the highly simplified mapping to an effective hard sphere model, which will be of different accuracy for chemically different classes of glass forming molecules. The local structure errors it incurs enter via the NLE dynamic free energy.

Concerning simulation tests of the core ideas underlying ECNLE theory, an analysis of the particle displacement field associated with the alpha event would be valuable, as done experimentally using confocal microscopy for glassy hard sphere colloidal suspensions [75]. There has been some effort in this direction in the precursor regime accessible on the computer, e.g., the democratic cluster [76] and metabasin [77] analyses. More work is required to search for the long range elastic distortion field that underpins the collective physics, although a very recent simulations [78] seem qualitatively consistent with our view.

Of course, the theory makes many approximations, of both a molecular model nature and with regards to the statistical mechanics. These include ignoring the explicit effects of nonspherical molecular shape and its consequences on packing, rotational versus translational motion and their coupling, possible anisotropy of the long range elastic strain field, and individual effects of repulsive versus attractive forces on structure and dynamics. Some of these might be explicitly addressed by building on our recent progress for nonspherical colloids using NLE theory [79-83]. Accounting for these factors at a microscopic level will "break" the quasi-universal nature of the present formulation based on effective hard spheres. In this regard, we note that for real molecules one expects the microscopic jump length, $\Delta r$, depends on non-universal details, and both the absolute magnitude and range of fragilities predicted by ECNLE theory are sensitive to this local dynamical quantity.

There are also the questions of dynamic heterogeneity and decoupling. Temporal heterogeneity is present at the cage level (e.g., Poissonian distribution of hopping times [84]), but will enter in a different, yet to be determined manner at the long range elastic distortion level. However, on general grounds we know that the local softening that occurs after the first relaxation event will reduce the collective elastic barrier in its spatial vicinity. This may then "facilitate" faster subsequent alpha events, in the spirit of the central ansatz of kinetic constraint models [18].

Despite the relative simplicity of the present theory, its predictive molecular character provides exciting opportunities to address other open problems of fundamental and materials science importance. For example, how does polymer conformation and connectivity determine the chain length dependence of the alpha time [4,5,85], and the uniquely polymeric "decoupling" of macromolecular versus segmental relaxation in melts [4,5,86, 87]? How does thin film confinement and surface/interface effects modify the alpha process and $T_g$? Is there a relationship between our dynamic approach and configurational and/or excess vibrational entropy [88] ideas? All these problems are presently under study.

**Acknowledgements**

We acknowledge support from the Division of Materials Science and Engineering, U.S. Department of Energy, Office of Basic Energy Sciences via ORNL. Discussions and/or correspondence with Jeppe Dyre, Mark Ediger, Alexei Sokolov, Matthieu Wyart, and Ernst Rossler are gratefully acknowledged.

---------------------------------

[1] M. Beiner, H. Huth, and K. Schröter, J. Non-Cryst. Sol., **279**, 126 (2001).
[2] G. Biroli, and J-P. Bouchaud, in *Structural Glasses and Supercooled Liquids:Theory, Experiment and Applications*, edited by P.G.Wolynes,V.Lubchenko eds.,Wiley & Sons, 2012) pp 34-113.
[3] C. A. Angell, K. L. Ngai, G. B. McKenna, P.F. McMillan, and S. W. Martin, J. Appl. Phys. **88**, 3113 (2000).
[4] C. M. Roland, Soft Matt., **4**, 2316 (2008).
[5] C.M.Roland, Macromolecules **43**, 7875 (2010).
[6] Paper I: S.Mirigian and K.S.Schweizer, J.Chem.Phys., preceding article.
[7] J-P. Hansen, I. R. McDonald, *Theory of Simple Liquids* (Academic Press, London 1986).
[8] T.S.Ingebrigtsen, T.B.Schroeder and J.C.Dyre, Phys.Rev.X, **2**, 011011 (2012).
[9] U. R. Pedersen, N. P. Bailey, T. B. Schrøder, and J. C. Dyre, Phys. Rev. Lett., **100,** 015701 (2008).
[10] U. R. Pedersen, T. Christensen, T. B. Schrøder, J. C. Dyre, Phys. Rev. E, **77**, 011201 (2008).
[11] J. C. Dyre, T. Christensen, N. B. Olsen, J. Non-Cryst. Solids, **352** 4635 (2006).
[12] J.C. Dyre, Rev. Mod. Phys., **78**, 953 (2006).
[13] L. Hong, P. D. Gujrati, V. N. Novikov, and A. P. Sokolov, J. Chem. Phys., **131**, 194511 (2009).
[14] L. Hong, V. N. Novikov, and A. P. Sokolov, Phys. Rev. E, **83**, 061508 (2011).
[15] W. Gotze, *Complex Dynamics of Glass-Forming Liquids: A Mode-Coupling Theory*, Oxford University Press, 2009.
[16] G. Adam and J. H. Gibbs, J. Chem. Phys., **43**, 139 (1965).
[17] V. Lubchenko and P. G. Wolynes, Ann. Rev. Phys. Chem., **58**, 235 (2007).
[18] D. Chandler and J. P. Garrahan, Ann. Rev. of Phys. Chem., **61**, 191 (2010).
[19] G.Tarjus, D.Kivelson and P.Viot, J.Phys.Condens.Matter, **12**, 6497 (2000).
[20] G. Tarjus and D. Kivelson, in *Jamming and Rheology: Constrained Dynamics on Microscopic and Macroscopic Scales*, edited by S. Edwards, A. Liu, and S. Nagel(Taylor Francis, New York 2001), pp 20-37.
[22] N. Petzold, B. Schmidtke, R. Kahlau, D. Bock, R. Meier, B. Micko, D. Kruk, and E. A. Rössler, J. Chem. Phys., **138**, 12A510 (2013).

[23] E. G. D. Cohen, R. Verberg, and I. M. de Schepper, Physica A, **251**, 251 (1998).
[24] R. Verberg, I. M. de Schepper, and E. G. D. Cohen, Phys. Rev. E, **55**, 3143 (1997).
[25] V. N. Novikov and E. A. Rössler, Polymer, **54**, 6987 (2013).
[26] H. A. Kramers, Physica, **7**(4), 284 (1940).
[27] P. Hänggi, P. Talkner, M. Borkovec, Rev. Mod. Phys., **62**(2), 251 (1990).
[28] E. J. Saltzman and K. S. Schweizer, J. Phys.: Condens. Matt., **19**, 205123 (2007).
[29] T. Kanaya, I. Tsukushi, K. Kaji, B. J. Gabrys, S. M. Bennington, and H. Furuya, Phys. Rev. B, **64**, 144202 (2001).
[30] T.Kanaya and K.Kaji, Adv.Polym.Sci., 154, 87 (2001).
[31] G. Parisi and F. Zamponi, Rev. Mod. Phys, **82**, 789 (2010).
[32] S.Torquato and F.H.Stillinger, Rev.Mod.Phys. **82**, 2633 (2010).
[33] K. S. Schweizer, J. Chem. Phys., **127**, 164506 (2007).
[34] C.Levelut, A.Faivre, R.LeParc, B.Champagnon, J.L.Hazemann and J.P.Simon, Phys.Rev.B, **72**, 224201 (2005).
[35] I. Saika-Voivod, F. Sciortino, and P. H. Poole, Phys. Rev. E, **63**, 011202(2000).
[36] J.Horbach and W.Kob, Phys.Rev.B, **60**, 3169 (1999).
[37] N.Kuzuu, K.Nagai, M.Tanaka, and Y.Tamai, Jap. J. App.Phys. **44**, 8086 (2005).
[38] J. Zhao, S. L. Simon, and G. B. McKenna, Nat. Comm., **4**, 1783 (2009).
[39] D. Kivelson, G. Tarjus, X. Zhao, and S. A. Kivelson, Phys. Rev. E, **53**(1), 751 (1996).
[40] B. Schmidtke, N. Petzold, R. Kahlau, M. Hofmann, and E. A. Rössler, Phys. Rev. E, **86**, 041507 (2012).
[41] M. O. McLinden and J. D. Splett, J. Res. Natl. Inst. Stand. Technol., **113**, 29 (2008).
[42] J. Opdycke, J. P. Dawson, R. K. Clark, M. Dutton, J. J. Ewing, and H. H. Schmidt, J. Phys. Chem., **68**(9), 2385 (1964).
[43] M. Naoki and S. Koeda, J. Phys. Chem., **93**, 948 (1989).
[44] L. Comez, S. Corezzi, D. Fioretto, H. Kriegs, A. Best, and W. Steffen, Phys. Rev. E, **70**, 011504 (2004).
[45] T. F. Sun, J. A. Schouten, and S. N. Biswas, Int. J. Thermophys., **12**(2), 381 (1991).
[46]*Handbook of Chemistry and Physics, 94th Ed.*, (46 Press, Boca Raton 2013), p. 6-160.
[47] M. Naoki, K. Ujita, and S. Kashima, J. Phys. Chem. **97**, 12356 (1993).
[48] N. B. Vargaftik, V. P. Kozhevnikov, and V. A. Alekseev, J. Eng. Phys., **35**(6), 1415 (1978).
[49] S. C. Jain and V. S. Nanda, J. Phys. C: Sol. State Phys., **4**, 3045 (1971).
[50] W. B. Streett, L. S. Sagan, and L. A. K. Stavely, J. Chem. Thermo., **5**(5), 633 (1973).
[51] E. Rössler, U. Warschewske, P. Eiermann, A. P. Sokolov, and D. Quitmann, J. of Non-Cryst. Sol., **172-174**, 113 (1994).
[52] R. Richert, K. Duwuri, and L-T. Duong, J. Chem. Phys., **118**, 1828 (2003).
[53] N. Menon, K. P. O'Brien, P. K. Dixon, L. Wu, S. R. Nagel, B. D. Williams, and J. P. Carini, J. of Non-Crystalline Solids, **141**, p61-65(1992).
[54] A. DöB, G. Hinze, B. Schiener, J. Hemberger, and R. Böhmer, J. Chem. Phys., **107**, 1740(1997)
[55] C. Alba, L. E. Busse, D. J. List, and C. A. Angell, J. Chem. Phys., **92**, 617 (1990).
[56] Q. Qin, G. B. McKenna, J. Non-Cryst. Sol. **352**, 2977 (2006).
[57] R. Böhmer, K. L. Ngai, C. A. Angell, and D. J. Plazek, J. Chem. Phys., **99**(5), 4201 (1993).
[58] C. Kliber, T. Hecksher, T. Pezeril, D. H. Torchinsky, J. C. Dyre, and K. A. Nelson, J. Chem. Phys., **138**, 12A544 (2013).
[59] J. C. Dyre, J. Non-Cryst. Solids, **235-237**, 142 (1998).
[60] C. M. Roland, S. Hensel-Bielowka, M. Paluch, and R. Casalini, Rep. Prog. Phys., **68**, 1405 (2005).
[61] A. Patkowski, M. Matos Lopes, and E. W. Fischer, J. Chem. Phys., **119**(3), 1579 (2003)
[62] C.Crauste-Thibierge, C.Brun, F.Ladieu, D.L'Hote, G.Biroli and J.P.Bouchaud, Phys.Rev.Lett. **104**, 165703 (2010).
[63] Th. Bauer, P. Lunkenheimer, and A. Loidl, Phys. Rev. Lett., **111**
[64] E.W.Fischer, PhysicaA, **210**, 183 (1993).
[65] D. Turnbull and M. H. Cohen, J. Chem. Phys., **52**(6), 3038 (1970).
[67] F. Stickel, E. W. Fischer, and R. Richert, J. Chem. Phys., **102**(15), 6251 (1995).
[68] K. L. Ngai, J. Non-Cryst. Sol., **275**, 7 (2000).
[70] Y. S. Elmatad, D. Chandler, and J. P. Garrahan, J. Phys. Chem. B, **113**, 5563 (2009).
[71] Y. S. Elmatad, D. Chandler, and J. P. Garrahan, J. Phys. Chem. B, **114**, 17113 (2010).
[72] S.L.Chan and S.R.Elliott, J.Phys.Condensed Matter **3**, 1269 (1991).
[73] C.A.Angell, J. Res. Natl. Inst. Stand. Technol., **102**, 171 (1997).
[74] H.Tanaka, Phys. Rev. Lett. **90**, 055701 (2003)
[75] P. Schall, D. A. Weitz, and F. Spaepen, Science, **318**, 1895 (2007).
[76] G.A.Appignanesi, J.A.Rodriguez-Fris, R.A.Montani and W.Kob, Phys.Rev.Lett. **96**, 057801 (2006).
[77] B.Doliwa and A.Heuer, Phys.Rev.E **67**, 031506 (2003).
[78] J. Chattoraj and A. Lemaître, Phys. Rev. Lett., **111**, 066001 (2013).
[79] G. Yatsenko and K. S. Schweizer, J. Chem. Phys., **126**, 014505 (2007).
[80] D. C. Viehman and K. S. Schweizer, Phys. Rev. E, **78**, 051404 (2008).
[81] M.Tripathy and K.S.Schweizer, J.Chem. Phys., **130**, 244906 (2009).
[82] J.Yang and K.S.Schweizer, J. Chem. Phys., **134**, 204908(2011).
[83] R. Zhang and K. S. Schweizer, Phys. Rev. E, **80**, 011502 (2009).
[84] E. J. Saltzman and K. S. Schweizer, J. Chem. Phys., **125**, 044509 (2006).
[85] J. Hintermeyer, A. Hermann, R. Kahlau, C. Goiceannu and E.A.Rossler, Macromolecules **41**, 9335 (2008).

[86] K.Ding and A.P.Sokolov, Macromolecules **39**, 3322 (2006)

[87] A. P. Sokolov and K. S. Schweizer, Phys. Rev. Lett., **102**, 248301 (2009).

[88] M. Wyart, Phys. Rev. Lett. **104**, 095901 (2010).

| | $N_s$ | A | B | $T_g$ (th,expt) | m (th,expt) | $T_A$ | $T_{A,eff}$, $T_X$ | $T_0$ | $T_B$, T' | $T_R$, $T^*$ | $T_{VFT}$ | $T_K$, $T_K$(alt) |
|---|---|---|---|---|---|---|---|---|---|---|---|---|
| Toluene[41] | 7 | 1.23 | 1158 | 166, 126[55] | 90, 115[#][55] | 321 | 230, 238 | 221 | 211, 204 | 185, 175 | 128 | 139, 116 |
| Biphenyl[42] | 12 | 1.34 | 1400 | 245, -- | 96, -- | 445 | 330, 340 | 319 | 306, 296 | 268, 257 | 192 | 207, 176 |
| OTP[43] | 18 | 0.43 | 1068 | 267, 246[56] | 82, 81[56] | 558 | 379, 394 | 365 | 346, 333 | 305, 281 | 201 | 220, 177 |
| TNB | 36 | 0.43 | 1068 | 362, 344[56] | 84, 86[56] | 721 | 504, 523 | 486 | 462, 446 | 406, 379 | 276 | 300, 246 |
| Salol[44] | 16 | 0.51 | 1104 | 257, 218[57] | 83, 73[57] | 532 | 364, 379 | 350 | 333, 320 | 293, 271 | 195 | 212, 172 |
| Ethanol[45] | 3 | 0.18 | 863 | 96, 92.5[56] | 79, 55[56] | 226 | 144, 150 | 137 | 129, 124 | 115, 102 | 71 | 78, 60 |
| Glycerol[46] | 6 | -1.3 | 992 | 203, 190[56] | 59, 53[56] | 785 | 352, 376 | 329 | 299, 283 | 280, 219 | 139 | 150, 99 |
| Sorbitol[47] | 12 | -1.33 | 1104 | 362, 266[56] | 53, 93[56] | 2288 | 697, 760 | 642 | 568, 535 | 560, 396 | 238 | 250, 145 |
| Sorbitol*[47] | 12 | -- | -- | 368, 266[56] | 62, 93[56] | 656 | 502, 516 | 589 | 465, 456 | 387, 393 | 274 | 286, 242 |
| Cesium[48] | 1 | -0.52 | 1675 | 114, -- | 73, -- | 291 | 175, 184 | 167 | 156, 149 | 139, 122 | 83 | 91, 69 |
| Rubidium[48] | 1 | -0.67 | 2185 | 150, -- | 73, -- | 389 | 232, 244 | 221 | 206, 197 | 185, 161 | 109 | 120, 90 |
| Argon[49] | 1 | 4.57 | 783 | 40, -- | 97, -- | 72 | 53, 55 | 52 | 49, 48 | 43, 42 | 31 | 34, 28 |
| Xenon[50] | 1 | 2.81 | 1194 | 67, -- | 87, -- | 131 | 93, 96 | 89 | 85, 82 | 75, 70 | 51 | 55, 46 |

**Table 1: Key parameters of the mapping, theoretical characteristic temperatures (Kelvin), and fragilities.** The mapping parameters are dimensionless except for B, which is in units of K. See the text for definitions of the various temperature scales. Experimental numbers are reported where available for the glass transition temperature and fragility. Results based on two definitions of a Kauzman-like temperature are reported; $T_K$ corresponds to Eq. 46 and $T_K$(alt) corresponds to Eq. 45. References given next to the molecule names refer to sources for the equation of state data used to determine the parameters A and B. Two sets of results are shown for sorbitol based on Eq. 8 and Eq. 2 (indicated by *)

[#]A wide range of fragilities are reported for toluene in the literature, ranging from 59[56] to 115[55].

|  | $T_A$ | $T_{A,eff}$, $T_X$ | $T_0$ | $T_B$, $T'$ | $T_R$, $T^*$ | $T_{VFT}$, $\overline{T}_K$ |
|---|---|---|---|---|---|---|
| Toluene | 1.93 | 1.38, 1.43 | 1.33 | 1.27, 1.23 | 1.11, 1.05 | 0.77, 0.77 |
| Biphenyl | 1.81 | 1.34, 1.39 | 1.3 | 1.25, 1.21 | 1.09, 1.05 | 0.78, 0.78 |
| OTP | 2.09 | 1.42, 1.48 (1.43) | 1.37 (1.39) | 1.3, 1.25 | 1.14, 1.05 | 0.75, 0.74 |
| TNB | 1.99 | 1.39, 1.44 | 1.34 (1.47) | 1.28, 1.23 | 1.12, 1.05 | 0.76, 0.75 |
| Salol | 2.06 | 1.42, 1.47 (1.35) | 1.36 (1.4) | 1.29, 1.24 | 1.14, 1.05 | 0.76, 0.75 |
| Ethanol | 2.37 | 1.5, 1.57 | 1.43 | 1.35, 1.29 | 1.2, 1.07 | 0.74, 0.72 |
| Glycerol | 3.88 | 1.74, 1.86 (1.68) | 1.62 | 1.48, 1.4 | 1.38, 1.08 | 0.69, 0.62 |
| Sorbitol | 6.32 | 1.93, 2.1 | 1.77 | 1.57, 1.48 | 1.55, 1.09 | 0.66, 0.55 |
| Sorbitol* | 1.78 | 1.36, 1.4 | 1.6 | 1.26, 1.24 | 1.05, 1.07 | 0.74, 0.72 |
| Rubidium | 2.6 | 1.55, 1.63 | 1.47 | 1.38, 1.31 | 1.23, 1.07 | 0.73, 0.7 |
| Cesium | 2.55 | 1.53, 1.61 | 1.46 | 1.37, 1.31 | 1.22, 1.07 | 0.73, 0.7 |
| Argon | 1.82 | 1.35, 1.39 | 1.3 | 1.25, 1.21 | 1.09, 1.05 | 0.78, 0.78 |
| Xenon | 1.97 | 1.39, 1.44 | 1.34 | 1.28, 1.23 | 1.12, 1.05 | 0.77, 0.76 |

**Table 2: Characteristic temperatures normalized by the theoretical $T_g$.** $\overline{T}_K$ represents the average of $T_K$ and $T_K$(alt), normalized by $T_g$. Where available, comparable numbers for $T_{A,eff}$[39] and for $T_0$[70] are reported in parentheses underneath our calculated numbers.

|  | $J/k_B T_0$ | $E_A/k_B$ (K) | $E_A/k_B T_g$ | $E_A/k_B T_{A,eff}$ | $\log(\tau_\alpha(T_{A,eff}))$ | $\log(\tau_\alpha(T_0))$ | $\log(\tau_\alpha(T_R))$ |
|---|---|---|---|---|---|---|---|
| Toluene | 9.99 | 1574 (1440) | 9.5 (12.3) | 6.9 | -10.18 | -9.79 | -5.04 |
| Biphenyl | 10.98 | 2470 | 10.1 | 7.5 | -10.1 | -9.71 | -4.31 |
| OTP | 8.97 (7.7-8.6) | 2357 (2441) | 8.8 (10) | 6.2 | -9.97 | -9.59 (-8.9) | -5.49 |
| TNB | 9.52 (7.1) | 3310 (3232) | 9.1 (9.4) | 6.6 | -9.8 | -9.41 (-9.2) | -4.96 |
| Salol | 9.1 (8.1-9.1) | 2294 (2104) | 8.9 (9.6) | 6.3 | -10.03 | -9.64 (-8.5) | -5.45 |
| Ethanol | 7.87 | 792 | 8.3 | 5.5 | -10.87 | -10.48 | -7.11 |
| Glycerol | 5.33 (4.1) | 1395 (2271) | 6.9 (12.1) | 4.0 | -10.33 | -9.94 (-7.7) | -8.25 |
| Sorbitol | 4.3 | 2380 | 6.6 | 3.4 | -10.3 | -9.91 | -8.86 |
| Sorbitol* | 6.14 | 5089 | 13.8 | 7.75 | -10.3 | -11.13 | -1.04 |
| Rubidium | 7.15 | 1175 | 7.8 | 5.1 | -10.65 | -10.26 | -7.36 |
| Cesium | 7.26 | 898 | 7.9 | 5.1 | -10.46 | -10.07 | -7.1 |
| Argon | 10.94 | 399 | 10.1 | 7.5 | -10.24 | -9.85 | -4.47 |
| Xenon | 9.67 | 616 | 9.3 | 6.7 | -10.03 | -9.64 | -5.09 |

**Table 3: Analysis of our theoretical calculations as data in the context of phenomenological models.** The left four columns in the table show relevant energy scales, while the right columns show relevant timescales. The parabolic law parameters J and $T_0$ correspond to Eq. 43; $E_A$ is the apparent Arrhenius energy, which describes our calculations up to a temperature $T_{A,eff}$; the temperature $T_R$ is where the magnitude of the two barriers are equal per the experimental analysis of Rössler and coworkers[40]. See text for details. Where available, the empirically deduced values from experimental fits[40,70] are reported in parentheses under our calculation.